\documentclass[table, xcdraw, sigconf]{acmart}
\acmConference[ESEC/FSE 2018]{12th Joint Meeting of the European Software Engineering Conference and the ACM SIGSOFT Symposium on the Foundations of Software Engineering}{4--9 November, 2018}{Lake Buena Vista, Florida}
\usepackage{lipsum}
\usepackage{graphicx}
\usepackage{booktabs} 
\usepackage{url}
\usepackage{subfig}
\usepackage{balance}
\usepackage{listings}
\usepackage[linesnumbered,ruled,vlined]{algorithm2e}
\usepackage{enumitem}
\setitemize{noitemsep,topsep=0pt,parsep=0pt,partopsep=0pt}

\usepackage[flushleft]{threeparttable}

\usepackage[tikz]{bclogo}
\newenvironment{RQ}[1]%
{\noindent\begin{minipage}[c]{\linewidth}%
\begin{bclogo}[couleur=gray!20,%
                arrondi=0.1,logo=\bctrombone,%
                ombre=true]{{\small ~#1}}}%
{\end{bclogo}\vspace{2mm}\end{minipage}}

\SetCommentSty{mycommfont}
\SetKwInOut{Parameter}{Parameter}
\SetKwInOut{Require}{Require Func}
\SetKwInOut{Input}{Input}
\SetKwInOut{Output}{Output}

\setcopyright{rightsretained}

\usepackage{enumitem}
\newcommand{\bi}{\begin{itemize}[leftmargin=0.4cm]}
\newcommand{\ei}{\end{itemize}}

\newcommand{\fig}[1]{Figure~\ref{fig:#1}}
\newcommand{\tabl}[1]{Table~\ref{tab:#1}}

\acmDOI{XX.YY/ZZ}

\acmISBN{ZZ-YY-24-ZZ/QQ/A}

\acmConference[FSE'18]{Florida}{Nov 2018}{} 
\acmYear{2018}
\copyrightyear{2018}

\acmPrice{15.00}

\acmSubmissionID{123-A12-B3}

\begin{document}
\title{Applications of Psychological Science for Actionable Analytics}

\author{Di Chen, Wei Fu, Rahul Krishna, Tim Menzies}
\affiliation{%
  \institution{North Carolina State University, USA}
  \city{Raleigh} 
  \state{NC} 
  \postcode{27606}
}
\email{{dchen20, wfu, rkrish11}@ncsu.edu,  tim.menzies@gmail.com}

\begin{abstract}
Actionable analytics are those that humans can understand, and operationalize.
What kind of data mining models generate such actionable analytics?
According to  
psychological scientists, humans understand models that most match their own internal models,
which they characterize as lists of ``heuristic'' (i.e., lists of very succinct rules).
One such heuristic rule generator is the  Fast-and-Frugal Trees (FFT) preferred by psychological scientists.
Despite their successful use in many applied domains, FFTs have not been applied in software
analytics.  
Accordingly, this paper assesses FFTs for software analytics. 

We find that FFTs are remarkably effective. Their models are very succinct (5 lines or less describing a binary decision tree).   These succinct models
outperform state-of-the-art defect
prediction algorithms defined by Ghortra et al. at ICSE'15.
Also, when we restrict training data to operational attributes (i.e., those attributes that are frequently changed by developers),
 FFTs perform much better than standard learners.

Our conclusions are two-fold. Firstly, 
 there is much that software analytics community could learn from psychological science.
Secondly, proponents of complex methods should always baseline those methods against simpler alternatives.
For example, FFTs could be used as a  standard baseline learner against which other software analytics tools are compared.
 
\end{abstract}

\keywords{Decision trees, heuristics, software analytics, psychological science,  
empirical studies, defect prediction}

%
%


\maketitle
\section{Introduction}
Data mining tools have been applied to many applications in Software Engineering (SE). For example, it has been used to estimate how long it would take to integrate new code into an existing project~\cite{czer11}, where defects are most likely to occur~\cite{ostrand04,Menzies2007a}, or how long will it take to develop a project~\cite{turhan11,koc11b}, etc. Large organizations like Microsoft routinely practice data-driven policy development where organizational policies are learned from an extensive analysis of large datasets~\cite{export:208800,theisen15}.

Despite these successes, there exists some drawbacks with current
software analytic tools. At a recent workshop on ``Actionable Analytics'' at ASE'15, business users
were very vocal in their
complaints about analytics~\cite{hihn15},  saying   
that  there are rarely  producible models that business users can understand or operationalize.

Accordingly, this paper explores methods for generating actionable analytics
for:
\bi
\item Software defect prediction;
\item Predicting close time for Github issues.
\ei
There are many  ways to define ``actionable'' but at the very least, we 
say that something is  actionable if  people can {\em read}
and  {\em use} the  models it generates.  Hence, for this paper, we   assume:

\vspace{2mm}
\centerline{{\bf {\em Actionable = Comprehensible  + Operational.}}}
\vspace{2mm}

\noindent
We show here that many algorithms  used in software analytics
generate models that are not actionable. Further, a data mining algorithm taken from
psychological science~\cite{czerlinski1999good,  gigerenzer1999good, 
martignon2003naive, 
brighton2006robust, 
martignon2008categorization, 
gigerenzer2008heuristics,  gigerenzer2011heuristic, neth2015heuristics}, called Fast-and-Frugal trees (FFTs\footnote{The reader might be aware that
FFT is also an acronym for ``Fast Fourier Transform''. Apparently, the psychological science community
was unaware of that acronym when they named this algorithm.}), are very
actionable. 

Note that  demanding that analytics   be actionable
also imposes certain restrictions
on (a)~the kinds of models that can be generated and (b)~the data used to build the models.
\bi
\item[(a)]
Drawing on psychological science, we  say  an
automatically generated model is {\em comprehensible} if:
\bi
\item
The model matches the  models
used internally by humans; i.e., it comprises small rules. 
\item
Further, for expert-level comprehension, the rules should quickly lead to decisions
(thus freeing up  memory for other tasks).
\ei
For more on this point, see Section~\ref{sec:compre}.
\item[(b)]
As to {\em operational}, we show  in
 the  historical log of software projects
 that only a few of the measurable project attributes
are often  changed
by developers.  
For a  data mining algorithm to be operational, it must generate effective
models even if  restricted to using just those  changed
attributes. 
\ei
Using three research questions, 
this paper tests if these
restrictions  damage our ability to build useful models.

{\bf RQ1: Do   FFTs models perform worse than the current state-of-the-art?} We  will find that:

\begin{RQ}{ For defect prediction,  FFTs   out-perform the state-of-art. }
When compared to state-of-the-art defect prediction algorithms surveyed by Ghotra et al.~\cite{ghotra2015revisiting}, FFTs are more effective
(where ``effective'' is measured in terms of  a recall/false alarm metric or    the $P_{\mathit{opt}}$  metric  
defined
 in \S~\ref{sect:eval}).
\end{RQ}

{\bf RQ2: Are FFTs more operational than the current state-of-the-art?} This research question tests what happens when we learn from less data; i.e., if we demand our models {\em avoid} using attributes that are
rarely changed by developers. We show that:

\begin{RQ}{When learning from less data, FFTs performance is   stabler than some other learners.}
When data is restricted to attributes that developers often
change, then FFTs performance is only slightly changed
while the performance of some other learners, can vary by alarmingly large amounts.
\end{RQ}
 
\noindent The observed superior performance of FFT raises the question:

{\bf RQ3: Why do FFTs work so well?} Our answer to this question will be somewhat technical but, in summary we will say:
 
\begin{RQ}{FFTs match the structure of SE data}
SE data divides into a few    regions with  very different properties and
FFTs are good way to explore such data spaces.
\end{RQ}


  
\noindent
In summary, the contributions of this paper are:
\bi
\item
A novel inter-disciplinary contribution of the application of psychological science to software
analytics.
\item
A cautionary tale that, for software analytics, {\em more} complex learners can perform  {\em worse}.
\item
A warning that many current results in software analytics make the, possibly unwarranted, assumption
that merely because an attribute is observable, that we should use those attributes in a model.
\item 
Three  tests for ``actionable analytics'': (a)~Does a data mining produce succinct models? 
(b)~Do those succinct models perform as well, or better, than more complex methods?
(c)~If the data mining algorithm is restricted to just the few attributes that developers
actually change, does the resulting model perform satisfactorily?
\item
A demonstration that the restraints demanding by actionable analytics (very simple models, access to less data) need not result in models with poor performance.
\item
A new, very simple baseline data mining method (FFTs) against which more complex methods
can be compared.
\item
A reproduction package containing all the data and algorithms of this paper, see \url{http://url_blinded_for_review}.
\ei

The rest of this paper is structured as follows.
In Section \ref{prelim}, we introduce the concepts of ``operational'' and ``comprehensible'' as the preliminaries. Our data, experimentation settings and evaluation measures will be described in Section \ref{methods}. In Section \ref{results}, we show our results and answer to research questions. Threats and validity of our work is given in Section \ref{threats}. In Section \ref{conclusions}, we conclude this paper with the following: 
\bi
\item
There is much the software analytics community could learn from psychological science.
\item
Proponents of complex methods should always baseline those methods against simpler alternatives. 
\ei
Finally, we discuss future work.


\section{Preliminaries} \label{prelim}
\subsection{ Operational}\label{sec:opt0}

This paper assumes that
for a  data mining algorithm to be operational, it must generate effective
models even if  restricted to using just those 
attributes which, in practice,
developers actually change. 
We have two reasons
for making that assumption.

 \begin{figure}[!t]
    \centering
    \includegraphics[width=.8\linewidth]{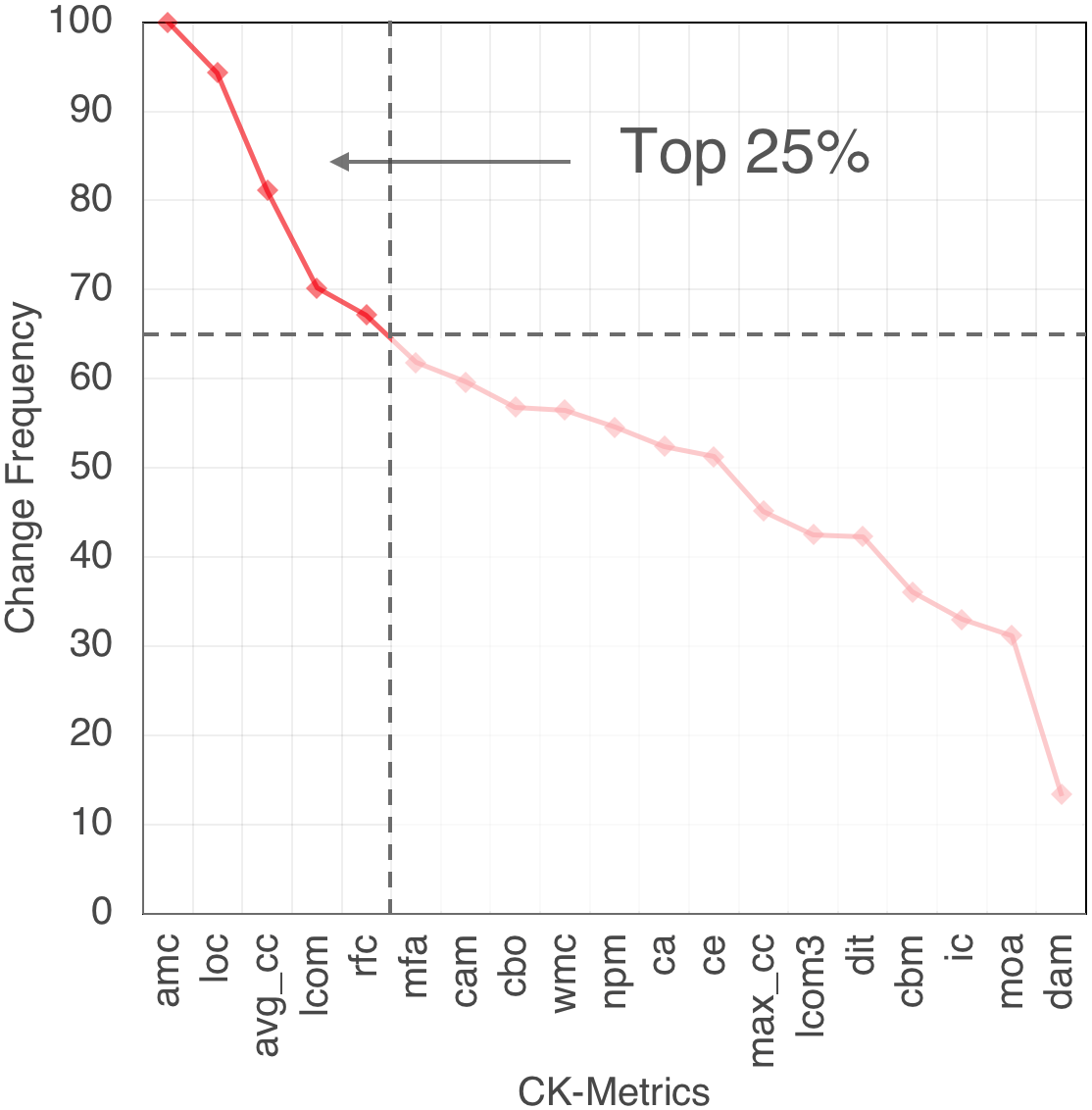}
    \caption{Only some metrics
      change between
      versions $i$ and $i+1$ of a software system.
      For definitions of the metrics
      on the x-axis, see \tabl{metrics}.
      To create this plot, we
      studied the  26  versions
      of the ten datasets
      in \tabl{data}.
      First we initialize $total=0$, then for all pairs of versions $i,i+1$ from the same data set,  we 
        (a)~incremented $\mathit{total}$ by one;
(b)~collected  the  distributions of metric $m$
seen in version $i$ and $i+1$ of the software;
(c)~ checked if  those two distributions were different;
and if so, (d)~added one to  $\mathit{changed_m}$.
Afterwards, the y-axis of this plot was
  computed using $100*\mathit{changed_m}/\mathit{total}$.}
    \label{fig:change_freq}
\end{figure}

Firstly, this definition of operational can make a model much more acceptable to developers.
If a model says that, say,  $x> 0.6$
  leads
to defective code then     
developers will  ask for guidance on how to
       reduce ``x''   (in order
       to reduce the chances of   
         defects).
       If we define ``operational'' 
       as per this article, then it is very simple matter
       to offer that developer numerous examples, from their own project's historical log, 
       of how ``x'' was changed.

Secondly, 
as shown in \fig{change_freq}
there exist attributes
that are usually not changed
from one version to the next.
\fig{change_freq}
  is important since,
as shown in our {\bf RQ2} results, when we restrict model construction
to just the 25\% most frequently changed attributes,
this can dramatically change the behavior
of some data mining algorithms (but not FFTs). 

Technical aside: in \fig{change_freq}, we defined
``changed''
 using the   A12 test ~\cite{vargha2000}  which
 declares  two distributions   
 different if they differ by more than a small effect. A recent ICSE'11 article~\cite{briand11} endorsed
 the use of A12 due to its non-parametric nature, it avoids any possibly
 incorrect Gaussian assumptions about the data.
  
 \begin{table*}
\scriptsize
    \renewcommand{\baselinestretch}{0.7}
    
    \caption{The C-K OO metrics studied in \fig{change_freq}. Note that the last line. \textit{`defect'}, denotes the dependent variable.}\label{tab:metrics}
    \begin{center}
    {
    \begin{tabular}{l|l|p{4.0in}}
    \rowcolor{lightgray}
    \hline
    Metric & Name & Description \\ \hline
    amc & average method complexity & Number of JAVA byte codes\\\hline
    avg\_cc & average McCabe & Average McCabe's cyclomatic complexity seen
    in class\\\hline
    ca & afferent couplings & How many other classes use the specific
    class. \\\hline
    cam & cohesion amongst classes & Summation of number of different
    types of method parameters in every method divided by a multiplication
    of number of different method parameter types in whole class and
    number of methods. \\\hline
    cbm &coupling between methods &  Total number of new/redefined methods
    to which all the inherited methods are coupled\\\hline
    cbo & coupling between objects & Increased when the methods of one
    class access services of another.\\\hline
    ce & efferent couplings & How many other classes is used by the
    specific class. \\\hline
    dam & data access & Ratio of  private (protected)
    attributes to   total   attributes\\\hline
    dit & depth of inheritance tree &  It's defined as the maximum length from the node to the root of the tree\\\hline
    ic & inheritance coupling &  Number of parent classes to which a given
    class is coupled (includes counts of methods and variables inherited)
    \\\hline
    lcom & lack of cohesion in methods &Number of pairs of methods that do
    not share a reference to an instance variable.\\\hline
    locm3 & another lack of cohesion measure & If $m,a$ are  the number of
    $methods,attributes$
    in a class number and $\mu(a)$  is the number of methods accessing an
    attribute, 
    then
    $lcom3=((\frac{1}{a} \sum_j^a \mu(a_j)) - m)/ (1-m)$.
    \\\hline
    loc & lines of code & Total lines of code in this file or package.\\\hline
    max\_cc & Maximum McCabe & maximum McCabe's cyclomatic complexity seen
    in class\\\hline
    mfa & functional abstraction & Number of methods inherited by a class
    plus number of methods accessible by member methods of the
    class\\\hline
    moa &  aggregation &  Count of the number of data declarations (class
    fields) whose types are user defined classes\\\hline
    noc &  number of children & Number of direct descendants (subclasses) for each class\\\hline
    npm & number of public methods & npm metric simply counts all the methods in a class that are declared as public. \\\hline
    rfc & response for a class &Number of  methods invoked in response to
    a message to the object.\\\hline
    wmc & weighted methods per class & A class with more member functions than its peers is considered to be more complex and therefore more error prone \\\hline
    
    defect & defect & Boolean: where defects found in post-release bug-tracking systems. \\\hline
    \end{tabular}
    }
    \end{center}
    
\end{table*}

\subsection{ Comprehensible}\label{sec:compre}
\subsubsection*{Why Demand Comprehensibility?}

This paper assumes that better data mining algorithms are better at explaining their models to humans.
But is that always the case?

The obvious counter-argument is that  if no human ever needs to understand our audited model, then it does not
need to be  comprehensible. For example, a neural net could control  the carburetor of an internal
combustion engine since that carburetor will never dispute the model or ask for clarification of any of
its reasoning.

On the other hand, if a model is to be used to persuade software engineers to change what they are doing, it needs to be  {\em comprehensible}   so humans can debate the merits of its conclusions. 
   Several researchers demand that software analytics models needs to be expressed in a simple way that is easy for software practitioners to interpret~\cite{menzies2014occam,  lipton2016mythos, dam2018explainable}.
According to Kim et al.~\cite{Kim2016},
 software analytics aim to
        obtain actionable insights
        from software artifacts that
        help practitioners accomplish tasks 
        related to software development, systems, 
        and users.          
   Other researchers~\cite{tan2016defining} argue that
        for software vendors, managers, developers and users, 
        such comprehensible  insights  are the core deliverable of software analytics. 
        Sawyer et al. comments that actionable insight is the key driver for businesses 
        to invest in data analytics initiatives~\cite{sawyer2013bi}.  
        Accordingly, much research focuses on the generation of   simple models, or make blackbox models more explainable, so that human engineers can understand and appropriately trust the decisions made by software analytics models~\cite{fu2017easy, abdollahi2016explainable}. 
      
If a model is not comprehensible, there are some explanation algorithms that might  mitigate that problem. For example:
\bi
\item
In  {\em secondary learning}, the examples given to a neural network are   used to train a rule-based learner and those learners could be said to ``explain'' the neural net~\cite{craven2014learning}. 
\item
In {\em contrast set learning} for instance-based reasoning,  data  is clustered  and users are shown the difference between a   few exemplars selected from   each cluster~\cite{krishna2015actionable}.
\ei
Such explanation facilities are post-processors to the original learning method. An alternative simpler approach would be to use learners that generate comprehensible models in the first place. 

The next section of this paper discusses one such alternate approach for creating simple comprehensible models.

\begin{table*}
\caption{Three example FFTs.}\label{tab:three}
 \begin{minipage}{.39\linewidth}
 {\footnotesize
\begin{verbatim}
  if      cob <= 4    then false     # 0 
  else if rfc > 32    then true      # 1   
  else if dam >  0    then true      # 1
  else if amc < 32.25 then true      # 1
  else false                         # 0
        \end{verbatim}}
\end{minipage}\begin{minipage}{.39\linewidth}
{\footnotesize
\begin{verbatim}
  if   	  cbo    <   4   then true # 1
  else if	max_cc <   3   then true # 1
  else if	wmc    <  10   then true # 1
  else if	rfc    <= 41.5 then true # 1
  else false                       # 0
  \end{verbatim}} \end{minipage}\begin{minipage}{.29\linewidth}
\vspace{-3mm}{\footnotesize \begin{verbatim}
  if	     dam > 0 then false # 0
  else if	noc > 0	then false # 0
  else if	wmc > 5	then false # 0
  else if	moa > 0	then false # 0
  else true                  # 1
\end{verbatim}}
\end{minipage}

\end{table*}

\subsubsection*{Theories of Expert Comprehension}
Psychological science argues that models comprising small rules are more comprehensible.
This section outlines that argument.

Larkin et al.~\cite{Larkin1335} characterize human expertise in terms of very small short term memory, or STM (used as a temporary scratch pad for current observation) and 
a very  large long term memory, or LTM.  
The LTM holds   separate tiny  rule fragments
that explore the contents
of STM to say ``when you see THIS, do THAT''.
When an LTM rule triggers, its
consequence can rewrite STM contents which,
in turn, can trigger other rules.

    Short term  memory is very  small, perhaps even as small as  four to seven items~\cite{Mi56,Co01}~\footnote{Recently,  Ma et al.~\cite{Ma14} used evidence from neuroscience and functional MRIs  to  argue  that STM capacity might be better measured using other factors than ``number of items''. But even they conceded that ``the concept of a limited (STM) has considerable explanatory power for behavioral data''.}.
Experts are experts, says Larkin et al.~\cite{Larkin1335} because the patterns in their  LTM
patterns dictate what to do, without needing to pause for reflection. Novices perform worse than experts,
says Larkin et al., when they fill  up  their STM with too many to-do's where they plan to pause and reflect on what to do next.  Since, experts post far fewer to-do's  in their STMs, they complete their tasks faster because (a) they are less encumbered by excessive reflection and (b) there is more space in their STM to reason about new information. 
While first proposed in 1981, this STM/LTM theory  still remains relevant~\cite{Ma14}. This theory can be used to explain both expert competency and incompetency in software
engineering tasks such as understanding code~\cite{Wi96}.

 Phillips et al.~\cite{phillips2017FFTrees} discuss how models containing tiny rule fragments can be  quickly comprehended by 
 doctors in emergency rooms making rapid  decisions; or by soldiers on guard  making snap decisions about whether to fire or not on a potential enemy; or by  stockbrokers making instant decisions about buying or selling stock. That is, according to this
 psychological science theory~\cite{czerlinski1999good, gigerenzer1999good, martignon2003naive, brighton2006robust, martignon2008categorization, gigerenzer2008heuristics, phillips2017FFTrees, gigerenzer2011heuristic,neth2015heuristics},
humans best understand a model:
\bi
\item
When they can ``fit'' it into their LTM; i.e., when that model comprises many small rule
fragments;
\item
Further, to have an expert-level comprehension of some
domain meaning having rules that can very quickly lead to decisions, without clogging up memory.
\ei

Psychological scientists have developed FFTs as one way
  to generate comprehensible models consisting of
   separate tiny rules~\cite{phillips2017FFTrees,gigerenzer2008heuristics,martignon2008categorization}.
      A FFT is a decision tree with exactly two branches extending
        from each node, where either one or both branches is an exit
        branch leading to a leaf~\cite{martignon2008categorization}. 
        That is to say, in an FFT,  every question posed by a node will
        trigger an immediate decision
        (so humans can read every leaf node
        as a separate rule).
        
        
        For example, \tabl{three} (at left)  is an  FFT generated from
        the log4j JAVA system of \tabl{data}.
        The goal of this tree is to classify a software module as ``defective=true'' or ``defective=false''.
         The four nodes in this FFT  reference  four static code attributes \emph{cbo,\ rfc,\ dam,\ amc}  (these metrics are defined in  
        \tabl{metrics}). 
        
FFTs are a binary classification algorithm. To apply such classifiers
to mulit-classes problems: (a)~build one  FFTs for each class
for classX or not classX; (b)~run all FFTs on the test example, then (c)~then select  conclusion  with most support (number of rows).

An FFT of depth $d$ has a choice of two ``exit policies'' at each level: the existing branch can select for the negation of the target (denoted ``0'') or the target (denoted ``1'').
The left-hand-side log4j tree in \tabl{three} is hence an 01110 tree since:
\bi
\item
The first level exits to the negation of the target: hence, ``0''.
\item
While the next tree levels
exit first to target; hence, ``111''.
\item
And the final line of the model exits
to the opposite of the penultimate line; hence, the final ``0''.
\ei
To build one FFT tree, select a maximum depth $d$, then follow the steps described in Table~\ref{tab:FFT-algo}

\begin{table}[!b]
\centering
\scriptsize
\caption{Some open-source JAVA systems. Used for training and testing showing different details for each. 
All   data   available on-line at http://tiny.cc/seacraft. }
\label{tab:data}
\resizebox{\columnwidth}{!}{%
\begin{tabular}{|l|l|l|l|l|l|}
\hline
\rowcolor[HTML]{C0C0C0} 
         & \multicolumn{2}{l|}{\cellcolor[HTML]{C0C0C0}Training} & \multicolumn{3}{l|}{\cellcolor[HTML]{C0C0C0}Testing} \\ \hline
\rowcolor[HTML]{C0C0C0} 
Data Set & Versions                         & Cases              & Versions        & Cases        & \% Defective        \\ \hline
jedit    & 3.2, 4.0, 4.1, 4.2               & 1257               & 4.3             & 492          & 2                   \\ \hline
ivy      & 1.1, 1.4                         & 352                & 2.0             & 352          & 11                  \\ \hline
camel    & 1.0, 1.2, 1.4                    & 1819               & 1.6             & 965          & 19                  \\ \hline
synapse  & 1.0, 1.1                         & 379                & 1.2             & 256          & 34                  \\ \hline
velocity & 1.4, 1.5                         & 410                & 1.6             & 229          & 34                  \\ \hline
lucene   & 2.0, 2.2                         & 442                & 2.4             & 340          & 59                  \\ \hline
poi      & 1.5, 2, 2.5                      & 936                & 3.0             & 442          & 64                  \\ \hline
xerces   & 1.0, 1.2, 1.3                    & 1055               & 1.4             & 588          & 74                  \\ \hline
log4j    & 1.0, 1.1                         & 244                & 1.2             & 205          & 92                  \\ \hline
xalan    & 2.4, 2.5, 2.6                    & 2411               & 2.7             & 909          & 99                  \\ \hline
\end{tabular}
}
\end{table}
\begin{table}[!b]
 
\caption{Steps for building FFTs}\label{tab:FFT-algo}
\begin{tabular}{|p{.95\linewidth}|}\hline
\rowcolor{gray!20}

(1) 	First discretize all attributes; e.g., split numerics on  median value.\\

(2) 	For each discretized range, find what rows it selects in the training data. 
		Using those rows, score each range using some user-supplied 
		$\mathit{score}$ function e.g., 
		recall, false alarm, or the $P_{\mathit{opt}}$ defined
		in \S\ref{sect:eval}.\\\rowcolor{gray!20}

(3) 	Divide the data on the best range.\\

(4) 	If the exit policy at this level is (0,1), 
		then exit to (false,true) 
		using the range that scores highest 
		assuming that the target class is (false,true), 
		respectively.\\\rowcolor{gray!20}

(5) 	If the current level is at $d$,  add one last exit node predicting the opposite to step 4. 
		Then terminate. \\

(6) 	Else, take the data selected by the non-exit range and go to step1
		to build the next level of the tree.\\\hline
\end{tabular}

\end{table}

 For trees of depth $d=4$, there are $2^4=16$ possible trees which we denoted 00001, 00010, 00101,... , 11110. Here,  the first four digits denote the 16 exit policies
 and the last digit denotes the last line of the model (which makes the opposite
 conclusion to the line above). For example:
 \bi
 \item
 A ``00001'' tree does it all it can to avoid the target class.
 Only after clearing away all the non-defective examples it can at levels one, two, three, four does it make a final ``true'' conclusion.  \tabl{three} (right) 
 shows the  log4j 00001 tree. Note that all the exits, except the last, are to ``false''.
 \item
  As to   ``11110'' trees, these   fixate on finding the target.
\tabl{three} (center)  shows
 the log4j 11110 tree. Note that all the exits, except the last, are to ``true''.
\ei
 During FFT training, we generate all $2^d$ trees then, using  the predicate $\mathit{score}$, select the best one (using the training data).
 This single best tree is then applied to the test data.
 
Following the advice of~\cite{phillips2017FFTrees}, for all the experiments of this paper, we use a depth    $d=4$. Note that FFTs of such small
depths are very succinct
(see above examples).  Many other data mining algorithms used in software analytics are far less
succinct and far less comprehensible (see \tabl{bad}).

\begin{table}[!t]
\caption{Comprehension issues with models generated by data mining algorithms used in software analytics.}\label{tab:bad}
\begin{tabular}{|p{.99\linewidth}|}\hline
\rowcolor{gray!20}
For very high dimensional data, there is some evidence that complex deep learning algorithms
    have  advantages for software engineering applications~\cite{yang2015deep, white2015toward,gu2016deep}. 
    However, since they do not readily support   explainability, they have been  criticizing   as 
    ``data mining alchemy''~\cite{DL2017alchemy}. \\

Support vector machines and principle component methods achieve their results after synthesizing new dimensions
which are totally unfamiliar to human users~\cite{Menzies2009ExplanationVP}.\\\rowcolor{gray!20}

Other methods that are heavily based on mathematics can be hard to explain to most
users. For example, in our experience, it is hard for (e.g.,) users to determine minimal changes to a project that mostly affect defect-proneness, just by browsing the internal frequency tables of a Naive Bayes classifier or the coefficients found via linear regression/logistic regression~\cite{Menzies2009ExplanationVP}.\\

When decision tree learners  are many pages long, they are  hard to browse and understand~\cite{friedl1997decision}.\\\rowcolor{gray!20}

Random forests are even harder to understand than decision trees since the problems of reading one tree are multiplied $N$ times, one for each member of the forest~\cite{liaw2002classification}. \\

Instance-based methods do not compress their training data; instead they produce  conclusions by  finding older exemplars closest to the new example. Hence, for such instance-based methods, it is hard to generalize and make a conclusion about what kind of future projects might be (e.g.,) most defective-prone~\cite{aha1991instance}.
\\\hline
\end{tabular}
\end{table}

 The value of models such as FFTs comprising many small rules has been extensively studied:
 \bi
 \item
             These    models use very few
             attributes from the data. Hence they tend to be robust against overfitting, especially on small and noisy data, and have been found to predict data at levels comparable with regression. See for example~\cite{martignon2008categorization, woike2017integrating,czerlinski1999good}.
        \item
        Other   work has  shown that these rule-based
        models can perform comparably well to more  complex models in a range of domains
        e.g., public health, medical risk management, performance science, etc.~\cite{jenny2013simple, laskey2014comparing, raab2015power}.
        \item
          Neth and Gigerenzer
          argue that such rule-bases  are tools that   work well under conditions of uncertainty~\cite{neth2015heuristics}. 
          \item
       Brighton showed that rule-based models can perform better than 
        complex nonlinear algorithms such as  neural networks, 
        exemplar models, and classification/regression trees~\cite{brighton2006robust}.
 \ei

\section{Methods} \label{methods}
The use of models comprising many small rules  has not been explored in the software analytics literature.
       This section describes the methods used by this paper to assess FFTs.
    \subsection{Data} \label{sec:data}

      \subsubsection{Defect Data:}
        To assess the FFTs, we perform our experiments using the publicly available SEACRAFT data~\cite{jureczko2010towards}, gathered by Jureczko et al. for object-oriented JAVA systems~\cite{jureczko2010towards}.
        The ``Jureczko'' data records the number of known defects for each class using a post-release defect tracking system. The classes are described in terms of nearly two dozen metrics such as number of children (noc), lines of code (loc), etc
        (see \tabl{metrics}). For details on the
        Jureczko data, see Table~\ref{tab:data}. 
        The nature of collected data and its relevance to defect prediction is discussed in greater detail by Madeyski \& Jureczko~\cite{madeyski2015process}.

        We selected these data sets since they have  at least three consecutive releases  (where release $i+1$ was built after release $i$). This is important for our experimental rig (see section 3.2).
        
        \begin{table}[!t]
\centering
\caption{Metrics used in issue lifetimes data}
\label{tab:issue_lifetime_metrics}
\resizebox{\columnwidth}{!}{%
\begin{tabular}{lll}
\textbf{Commit}                              & \textbf{Comment}                      & \textbf{Issue}                \\ \hline
\multicolumn{1}{l|}{nCommitsByActorsT}       & \multicolumn{1}{l|}{meanCommentSizeT} & issueCleanedBodyLen           \\
\multicolumn{1}{l|}{nCommitsByCreator}       & \multicolumn{1}{l|}{nComments}        & nIssuesByCreator              \\
\multicolumn{1}{l|}{nCommitsByUniqueActorsT} & \multicolumn{1}{l|}{}                 & nIssuesByCreatorClosed        \\
\multicolumn{1}{l|}{nCommitsInProject}       & \multicolumn{1}{l|}{}                 & nIssuesCreatedInProject       \\
\multicolumn{1}{l|}{nCommitsProjectT}        & \multicolumn{1}{l|}{}                 & nIssuesCreatedInProjectClosed \\
\multicolumn{1}{l|}{}                        & \multicolumn{1}{l|}{}                 & nIssuesCreatedProjectClosedT  \\
\multicolumn{1}{l|}{}                        & \multicolumn{1}{l|}{}                 & nIssuesCreatedProjectT        \\ \hline
\multicolumn{1}{l|}{Misc.}                   & \multicolumn{2}{l}{nActors, nLabels, nSubscribedByT}                 \\ \hline
\end{tabular}
}
\end{table}
          \subsubsection{Issue Lifetime Data:}
          This paper will conclude that FFTs
          are remarkable effective.  
        To check the external validity of that conclusion,
         we will apply FFT to another SE domain~\cite{rahul2018bellwhether,rees2017better}.
         Our Github issue lifetime data\footnote{https://doi.org/10.5281/zenodo.197111}
          consists of 8 projects used to study issue lifetimes.  In raw form, the data consisted of sets of JSON files for each repository, each file contained one type of data regarding the software repository (issues, commits, code contributors, changes to specific files as shown in Table \ref{tab:issue_lifetime_metrics}). 
        In order to extract data specific to issue lifetime, we did similar preprocessing and feature extraction on the raw datasets as suggested by \cite{rees2017better}.
        
        \subsection{Experimental Rig}
      For the defect prediction data, we use versions $i,j,k$ of the software systems in \tabl{data}.

         Using versions $i,j$,
        we track what attributes change by from version $i$ to $j$ (using the calculation shown in \fig{change_freq}). Then we build a model
        using {\em all} the attributes from version $j$ or just the top 25\% most changed attributes.
       Note that this implements our definition of ``operational'',
        as discussed in our introduction.
        
        After building a model, we use   the latest version $k$ for testing while the older versions for training.  In this way, we can assert that all our predictions
        are using past date to predict the future.

     For the issue lifetime data, we do not have access to multiple versions of the data. Hence, for this data we cannot perform the operational test.
     Hence, for that data we conduct a 5*10   cross-validation experiment that ensures that the train and test sets are different. For that cross-val, we divide the data into ten bins, then for each bin $b_i$ we train on $\mathit{data} - b_i$ then
     test on bin $b_i$.  To control for order effects
     (where the conclusions are altered by the order of the input examples)~\cite{agrawal2018wrong},
     this process is repeated five times, using different random orderings of the data.

    \subsection{Data Mining Algorithms}
\begin{table}[!t]
\centering
\small
\caption{For the purposes of predicting
software defects, Ghotra et al.~\cite{ghotra2015revisiting} found that many learners
have similar performance. Here are their four clusters of 32 data mining algorithms.
For our work, we selected learners  at random, one  from each cluster (see the \underline{\textbf{underlined}}
entries). }
\label{tab:learners}
\resizebox{\columnwidth}{!}{%
\begin{tabular}{|l|l|l|l|l|}
\hline
\rowcolor[HTML]{C0C0C0} 
\begin{tabular}[c]{@{}l@{}}Overall \\ Rank\end{tabular} & \begin{tabular}[c]{@{}l@{}}Classification \\ Technique\end{tabular}                                                               & \begin{tabular}[c]{@{}l@{}}Median \\ Rank\end{tabular} & \begin{tabular}[c]{@{}l@{}}Average \\ Rank\end{tabular} & \begin{tabular}[c]{@{}l@{}}Standard \\ Deviation\end{tabular} \\ \hline
1                                                       & \begin{tabular}[c]{@{}l@{}}Rsub+J48, \underline{\textbf{SL}}, Rsub+SL, \\ Bag+SL, LMT, RF+SL, \\ RF+J48, Bag+LMT, \\ Rsub+LMT, and RF+LMT\end{tabular} & 1.7                                                    & 1.63                                                    & 0.33                                                          \\ \hline
2                                                       & \begin{tabular}[c]{@{}l@{}}RBFs, Bag+J48, Ad+SL, \\ KNN, RF+NB, Ad+LMT, \\ \underline{\textbf{NB}}, Rsub+NB, and Bag+NB\end{tabular}                   & 2.8                                                    & 2.84                                                    & 0.41                                                          \\ \hline
3                                                       & \begin{tabular}[c]{@{}l@{}}Ripper, \underline{\textbf{EM}}, J48, Ad+NB, \\ Bag+SMO, Ad+J48, \\ Ad+SMO, and K-means\end{tabular}                        & 5.1                                                    & 5.13                                                    & 0.46                                                          \\ \hline
4                                                       & \begin{tabular}[c]{@{}l@{}}RF+SMO, Ridor, \underline{\textbf{SMO}}, \\ and Rsub+SMO\end{tabular}                                                       & 6.5                                                    & 6.45                                                    & 0.25                                                          \\ \hline
\end{tabular}
}
\end{table}
      
      The results shown below compare FFTs to 
      state of the art algorithms from software analytics.
      For a list of state-of-algorithms, we used 
       the ICSE'15 paper from Ghotra et al.~\cite{ghotra2015revisiting} which compared 32   classifiers for defect prediction. Their statistical analysis showed that the performance of these classifiers clustered into four groups shown in Table~\ref{tab:learners}.
        For our work, we selected one classifier at random from each of their clusters:  i.e., Simple Logistic (SL), Naive Bayes (NB), Expectation Maximization (EM), Sequential Minimal Optimization (SMO).
        
        Simple Logistic and Naive Bayes falls into the 1st and 2nd rankings layers. They are both statistical techniques that are based on a probability based model ~\cite{kotsiantis2007supervised}.
        These techniques are used to find patterns in datasets and build
        diverse predictive models~\cite{berson2004overview}. 
        Simple Logistic is a generalized linear regression
        model that uses a logit function.
        Naive Bayes is a probability-based technique
        that assumes that all of the predictors are independent of
        each other.

        Clustering techniques like EM divide the training data into small
        groups such that the similarity within groups is more than
        across the groups ~\cite{hammouda2000comparative}. 
        EM is a 
        clustering technique based on  cluster performance Expectation Maximization~\cite{fraley2007bayesian}
        (EM) technique, which automatically splits a dataset into an (approximately) optimal number of clusters~\cite{bettenburg2012think}.
        
        Support Vector Machines (SVMs) use a hyperplane to
        separate two classes (i.e., defective or not). 
        In this paper, following the results of Ghotra et al., we use the Sequential Minimal Optimization
        (SMO) SVM technique. SMO analytically solves the large
        Quadratic Programming (QP) optimization problem which
        occurs in SVM training by dividing the problem into a series
        of possible QP problems~\cite{zeng2008fast}.

    \subsection{Evaluation Measures}\label{sect:eval}

        Our rig assess learned models using an evaluation function
        called $\mathit{score}$. For FFTs,
        this function is called three  times:
        \bi
        \item Once to rank discretized ranges; 
        \item Then once again to select the best FFT out of the $2^d$ trees generated
        during training.
        \item Then finally, $\mathit{score}$ is used to score what happens when that best FFT is applied to the test data.
        \ei
        For all the other learners, {\em score} is applied on the test data.
        For this work, we use the two $\mathit{score}$ measures:
        {\em dis2heaven} and $P_{opt}$.

          Ideally, a perfect learner will have perfect recall (100\%) with no false alarms. \begin{equation} \label{eq:recall}
            \mathit{Recall} = \frac{\mathit{True Positive}}{\mathit{True Positive} + \mathit{False Negative}}
        \end{equation}
        \begin{equation} \label{eq:far}
            \mathit{FAR} = \frac{\mathit{False Positive}}{\mathit{False Positive} + \mathit{True Negative}}
        \end{equation}
        We combine these two into a ``distance to heaven'' measure called
        {\em dis2heaven} that reports how far a learner falls away from the ideal point of {\em Recall=1} and {\em FAR=0}:
        \begin{equation} \label{eq:d2h}
            \mathit{score}_1 = \mathit{dis2heaven} = \sqrt{\frac{{(1-\mathit{Recall})}^2 + {\mathit{FAR}}^2}{2}}
        \end{equation}

      \begin{figure}[!t]
            \centering
            \includegraphics[width=1.05\columnwidth]{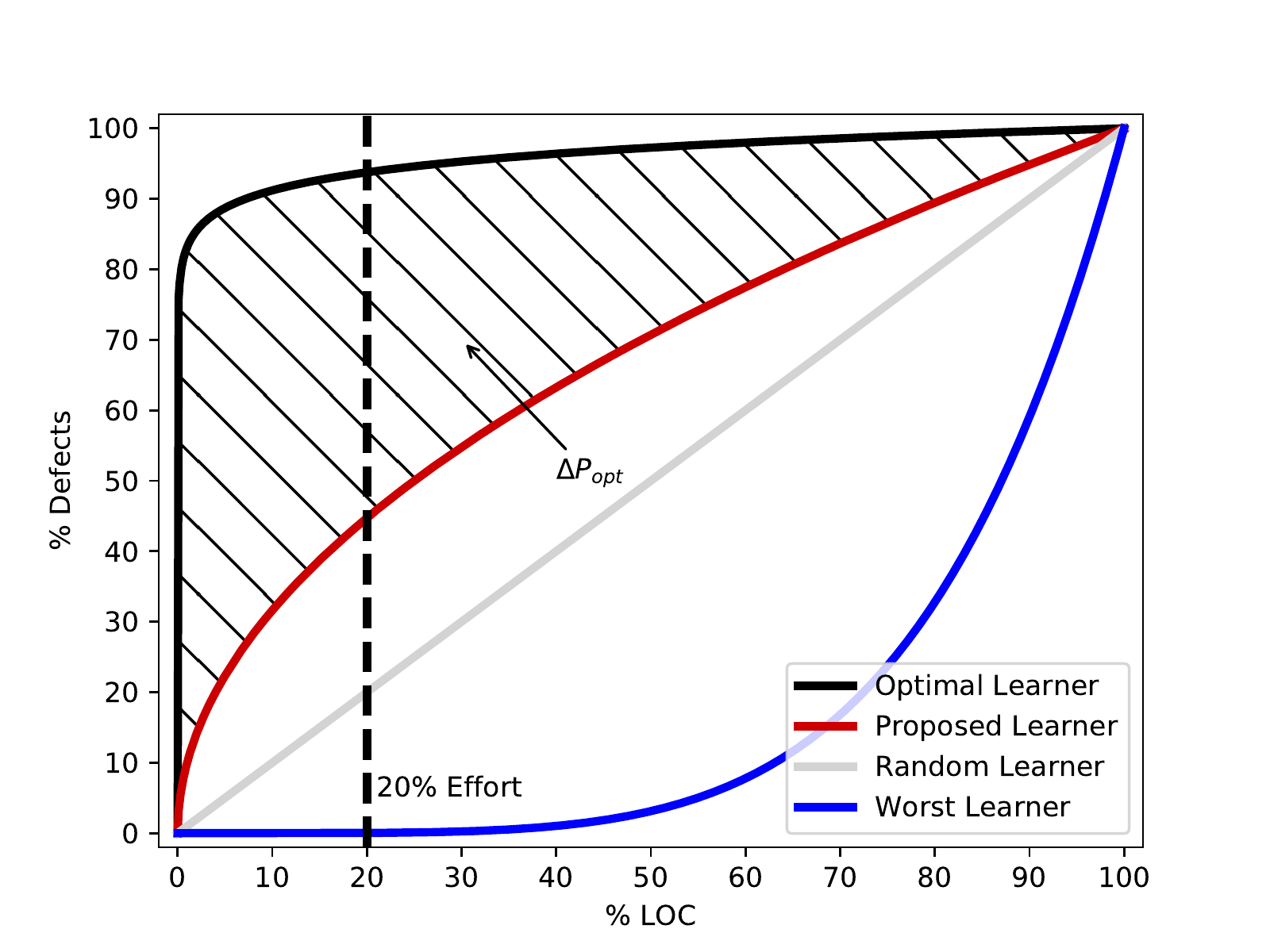}
            \caption{Effort-based cumulative lift chart~\cite{yang2016effort}.}
            \label{fig:p_opt_demo}
        \end{figure}    
        
As to $P_{opt}$,
        Ostrand et al. \cite{ostrand2005predicting} report that their quality predictors can find  20\% of the files contain on average 80\% of all defects in the project. Although there is nothing magical about the number 20\%, it has been used as a cutoff value to set the efforts required for the defect inspection when evaluating the defect learners ~\cite{kamei2013large, mende2010effort, monden2013assessing, yang2016effort}. 
        That is, $P_{opt}$ reports how many defects have been found after
        (a)~the code is sorted by the learner from ``most likely to be buggy''
        to ``least likely''; then (b)
        humans inspect 20\% of the code (measured in lines of code),
        where that code has , how many defects can be detected by the learner. 
        This measure is  widely used in defect prediction literature ~\cite{kamei2013large, menzies2007data, menzies2010defect, monden2013assessing, yang2016effort, zimmermann2007predicting}.
        
        $P_{\mathit{opt}}$ is defined as $1-\Delta_{\mathit{opt}}$ , where $\Delta_{\mathit{opt}}$ is the area between the effort cumulative lift charts of the optimal model and the prediction model (as shown in Figure\ref{fig:p_opt_demo}). 
        In this chart, the x-axis is  the percentage of required effort to inspect the code and the y-axis is the percentage of defects found in the selected code. In the optimal model, all the changes are sorted by the actual defect density in descending order, while for the predicted model, all the changes are sorted by the actual predicted value in descending order. According to Kamei et al. and Xu et al. ~\cite{kamei2013large, monden2013assessing, yang2016effort} $P_{\mathit{opt}}$ can be normalized as follows:
            \begin{equation} \label{eq:popt}
          \mathit{score}_2 =   P_{\mathit{opt}}(m) = 
            1 - \frac{S(\mathit{optimal}) - S(m)}{S(\mathit{optimal}) - S(\mathit{worst})}
            \end{equation}
        where $S(\mathit{optimal})$, $S(m)$ and 
        $S(\mathit{worst})$ represent the area of curve under the optimal model, predicted model, and worst model, respectively.
     This  worst model is built by sorting all the
        changes according to the actual defect density in ascending order.  
        
          Note that for our two {\em score} functions:
          \bi
          \item 
          For {\em dis2heaven}, the
          {\em lower} values are {\em better}.
          \item
          For $P_{opt}$, the
          {\em higher} values are {\em better}.
        \ei

      \begin{figure*} 
                \centering
                    \includegraphics[width=.85\columnwidth]{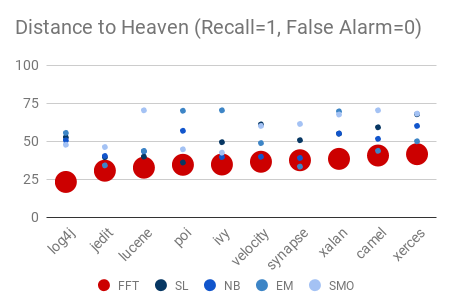} ~~~~~~~~~~~ \includegraphics[width=.85\columnwidth]{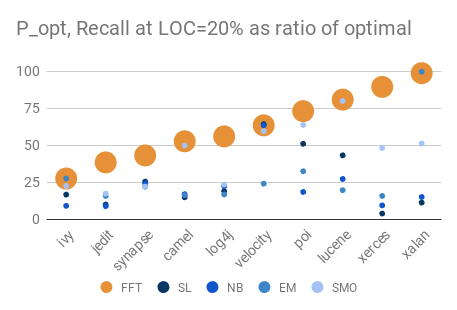}
                \caption{ 
                On the  left, in the   {\em dis2Heaven} results,
                  {\em less } is {\em better}.
                  On the right, in the $P_{\mathit{opt}}$ results,   {\em more}
                is {\em better}.
              On both sides, the
              FFTs results are better
              than those from   state-of-the-art   defect prediction algorithms (as defined by  Ghotra et al.~\cite{ghotra2015revisiting}).}
                \label{fig:cmp-dt2}

                \end{figure*}
                

\section{Results} \label{results}

    \subsection{RQ1: Do   FFTs models perform worse than the current state-of-the-art?}


         Figure~\ref{fig:cmp-dt2} compares the performance of FFT versus learners taken from
         Ghotra et al.  In this figure, datasets are sorted left right based on the FFT performance scores.
         With very few exceptions:
        \bi
        \item 
        FFT's {\em dis2heaven}'s results  
         {\em lower}, hence  {\em better}, than the other learners.
         \item
         FFT's $P_{\mathit{opt}}$ results
         are  much {\em higher}, hence
         {\em better}, than the other learners.
         \ei
          Therefore our answer to {\bf RQ1} is:
          
          \begin{RQ}{
For defect prediction,  FFTs   out-perform the state-of-art. }
When compared to state-of-the-art defect prediction algorithms surveyed by Ghotra et al., FFTs are more effective
(where ``effective'' is measured in terms of  a recall/false alarm metric or    $P_{\mathit{opt}}$).
\end{RQ}

  \begin{figure*}
          \label{fig:fft-compare}
          \includegraphics[width=\textwidth]{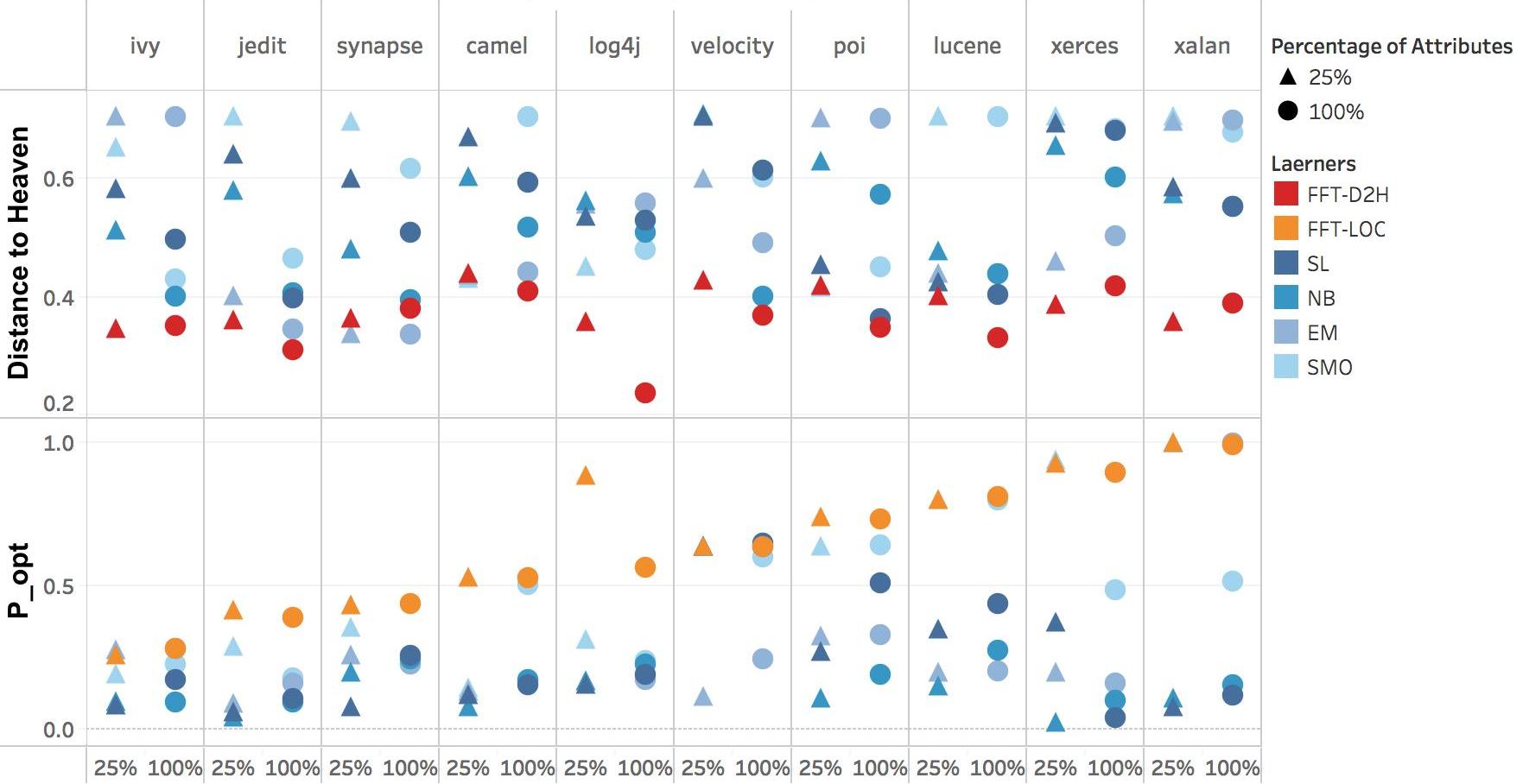}
          \caption{For each learner in \fig{cmp-dt2}, this plot shows the  difference between the  results obtains using the  top 25\% or  all (100\%) of attributes.
            For ($dist2heaven$,$P_{\mathit{opt}}$), values that are 
            ({\em lesser,greater})   (respectively)
            are {\em better}.
            Note that all the {\Large$\bullet$} 100\% results were also shown in \fig{cmp-dt2}.}\label{fig:compare}
        \end{figure*}
       
     \begin{figure}
        \centerline{\includegraphics[width=\linewidth]{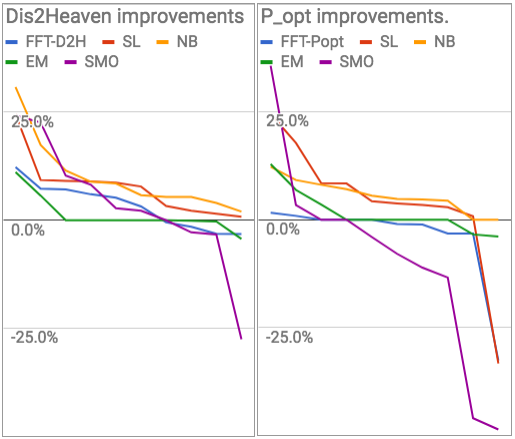}}
        
        \vspace{5mm}
        \caption{Deltas between results  25\% and 100\% of the data. Computed from \fig{compare}.
        Calculated such that {\em larger} values are {\em better};
        i.e., for 
        ({\em dist2heaven},
        $P_{\mathit{opt}}$)
        we report   (25\%-100\%, 100\%-25\%) since  (less, more) values are better (respectively). All values for each learner are sorted independently.}\label{fig:diffs}
        \end{figure}

    \subsection{RQ2: Are FFTs more operational than the current state-of-the-art?} 
Please recollect from before
that   a model is operational if its performance is not affected
after  {\em avoiding}
      attributes   that are rarely changed
    by developers.

    \fig{compare}
        compares
        model performance when we learn
    from all 100\% attributes or just the 25\% most
    changed attributes. For this study,
    these 25\% group (of most changed attributes)
    was computed separately for each data set. Note that:
    \bi
    \item The top row of \fig{compare}
    shows the {\em dis2heaven} results;
    \item The bottom row of \fig{compare}
    shows the $P_{\mathit{opt}}$
    results.
    \ei  
    \fig{diffs} reports the deltas in performance scores seen between using 25\% and 100\% of the data. These deltas are computed  such that {\em larger} values are {\em better};
        i.e., for 
        ({\em dist2heaven},
        $P_{\mathit{opt}}$)
        we report   (25\%-100\%, 100\%-25\%) since  (fewer, more) values are better (respectively).
    
    There are several key features for these results:
    \bi
    \item 
    The FFT's red dots for {\em dis2heaven} are {\em below} the rest; also,  FFT's orange dots for $P_{\mathit{opt}}$ are {\em above}
    the rest. This means that,   regardless of whether we use all attributes or just the most changed attributes, 
    the FFT results are nearly always better than the other methods.
    \item As seen in \fig{diffs}, the deltas between
    using all data and just some of the data is smallest for
    FFTs and EM (the instance-based clustering algorithm). In $P_{\mathit{opt}}$, those deltas are very small indeed (the FFT
    and EM results lie right on the y-axis for most of that plot).
    \item Also, see in \fig{diffs}, the deltas on the other learners
    can be highly  variable. While for the most part, using just the 25\% most
    changed attributes improves performance, SMO , SL and NB all have
    large negative results for at least some of the data sets. 
    \ei
    \noindent In summary, the learners studied here fall into three groups:
    \begin{enumerate}
    \item  Those that   exhibited a wide 
    performance variance   after  restricting the learning
    to just the frequently changed data (SL, NB, SMO),
   and those that are not (FFT, EM);
    \item  Those with best performance across
    the two performance measures studied here (FFT),
    and the rest (SL, NB, EM, SMO);
    \item Those that generate tiny models (FFT),
    and the rest (SL, NB, EM, SMO).
    \end{enumerate}
    Accordingly, FFT is the recommended learner since it  both performs well and is unaffected by issues such as whether or not the  data is restricted to just the most operational attributes.
    In summary:

\begin{RQ}{When learning from less data, FFTs performance is   stabler than some other learners.}
When data is restricted to attributes that developers often
change, then FFTs performance is only slightly changed
while the performance of some other learners, can vary by alarmingly large amounts.
\end{RQ}

\begin{table}
  \centering
  
  \footnotesize
  \caption{ Frequency heatmap of best exit polices seen for  FFT and defect prediction.}\label{tab:xcept}
  \vspace{5mm}
  \resizebox{\columnwidth}{!}{%
  \label{tab:fft-frequency}%
    \begin{tabular}{c|r|r|r|r|r}
    \rowcolor[rgb]{ .851,  .851,  .851} \multicolumn{1}{l|}{Best FFF} &       & \multicolumn{2}{c|}{\textbf{25\%  }} & \multicolumn{2}{c}{\textbf{100\%  }} \\
    
    \rowcolor[rgb]{ .851,  .851,  .851} \multicolumn{1}{p{5em}|}{\textbf{exit policy}} &       & \multicolumn{1}{p{2.25em}|}{D2H} & \multicolumn{1}{p{2.25em}|}{$P_{\mathit{opt}}$ } & \multicolumn{1}{p{2.25em}|}{D2H} & \multicolumn{1}{p{2.25em}}{$P_{\mathit{opt}}$} \\
    
    
    \rowcolor[rgb]{ .988,  .988,  1} 00001 & 0     & 0     & 0     & 0     & 0 \\
    \rowcolor[rgb]{ .988,  .988,  1} 00010 & 0     & 0     & 0     & 0     & 0 \\
    \rowcolor[rgb]{ .988,  .988,  1} 00101 & 0     & 0     & 0     & 0     & 0 \\
    \rowcolor[rgb]{ .988,  .988,  1} 00110 & 0     & 0     & 0     & 0     & 0 \\
    \rowcolor[rgb]{ .988,  .988,  1} 01001 & 0     & 0     & 0     & 0     & 0 \\
    \rowcolor[rgb]{ .988,  .988,  1} 01010 & 0     & 0     & 0     & 0     & 0 \\
    \rowcolor[rgb]{ .988,  .988,  1} 01101 & \cellcolor[rgb]{ .988,  .949,  .961}1 & 0     & 0     & 0     & \cellcolor[rgb]{ .988,  .949,  .961}1 \\
    \rowcolor[rgb]{ .988,  .988,  1} 01110 & 0     & 0     & 0     & 0     & 0 \\
    \rowcolor[rgb]{ .988,  .988,  1} 10001 & \cellcolor[rgb]{ .973,  .412,  .42}14 & \cellcolor[rgb]{ .984,  .741,  .753}6 & 0     & \cellcolor[rgb]{ .98,  .702,  .71}7 & \cellcolor[rgb]{ .988,  .949,  .961}1 \\
    \rowcolor[rgb]{ .988,  .988,  1} 10010 & \cellcolor[rgb]{ .980,  .659,  .671}8 & \cellcolor[rgb]{ .984,  .824,  .835}4 & \cellcolor[rgb]{ .988,  .906,  .918}2 & \cellcolor[rgb]{ .988,  .906,  .918}2     & 0 \\
    \rowcolor[rgb]{ .988,  .988,  1} 10101 & \cellcolor[rgb]{ .988,  .867,  .878}3 & 0     & \cellcolor[rgb]{ .988,  .949,  .961}1 & \cellcolor[rgb]{ .988,  .949,  .961}1 & \cellcolor[rgb]{ .988,  .949,  .961}1 \\
    \rowcolor[rgb]{ .988,  .988,  1} 10110 & \cellcolor[rgb]{ .984,  .784,  .796}5 & 0     & \cellcolor[rgb]{ .988,  .867,  .878}3 & 0     & \cellcolor[rgb]{ .988,  .906,  .918}2 \\
    \rowcolor[rgb]{ .988,  .988,  1} 11001 & 0     & 0     & 0     & 0     & 0 \\
    \rowcolor[rgb]{ .988,  .988,  1} 11010 & \cellcolor[rgb]{ .988,  .867,  .878}3 & 0     & \cellcolor[rgb]{ .988,  .949,  .961}1 & 0     & \cellcolor[rgb]{ .988,  .906,  .918}2 \\
    \rowcolor[rgb]{ .988,  .988,  1} 11101 & \cellcolor[rgb]{ .988,  .906,  .918}2 & 0     & 0     & 0     & \cellcolor[rgb]{ .988,  .906,  .918}2 \\
    \rowcolor[rgb]{ .988,  .988,  1} 11110 & \cellcolor[rgb]{ .984,  .824,  .835}4 & 0     & \cellcolor[rgb]{ .988,  .867,  .878}3 & 0     & \cellcolor[rgb]{ .988,  .949,  .961}1 \\\hline
   Totals &  40 & 10 & 10 & 10 & 10 
    \end{tabular}%
    }
    \vspace{5mm}
\end{table}%

    \subsection{RQ3: Why do FFTs work so well?}

 To  explain the success of FFTs, 
recall that during training, FFTs explores $2^d$ models, then selects the models whose
 exit policies achieves best performances (exit policies were introduced in Section ~\ref{sec:compre}). 
The exit policies selected by FFTs are like a trace of the reasoning
 jumping around the data.
 For example, a 11110 policy shows a model always jumping   towards sections of the data
 containing most defects. Also, a 00001 policy show another model trying to jump away from defects  until, in its last step, it does one final jump towards  defects.
 \tabl{xcept} shows what exit policies were seen in the experiments of the last section:
 \bi
 \item
 The 11110 policy was used sometimes. 
 \item
 A more common policy is 10001
 which shows a tree first jumping to some low hanging fruit (see the first ``1''),
 then jumping away from defects three times (see the next ``000'') before a final jump into defects (see the last ``1''). 
 \item
 That said, 
 while 10001 was most common, many other exit policies appear in \tabl{xcept}. For example,
 the $P_{\mathit{opt}}$  policies are particularly diverse. 
 \ei
\tabl{xcept}  suggests that software data is ``lumpy''; i.e., it divides
into a few separate   regions, each with different properties.  Further, the number and importance of the ``lumps'' is 
specific to the data set and the goal criteria.
In such a ``lumpy'' space,
a learning policy like FFT works well since its exit policies let a learner discover how to best
jump between the ``lumps''. Other learners  fail in this coarse-grained  lumpy space  when they:
\bi
\item
Divide the data too much; e.g. like RandomForests, which finely divide the data multiple times down the branches of the trees and across multiple trees;
\item
Fit some general model across all the different parts of the data; e.g. like simple logistic regression.
\ei
In summary, in answer to the question ``why do FFTs work so well'', we reply:

\begin{RQ}{FFTs match the structure of SE data}
 SE data divides into a few regions with very different properties and FFTs are good way to explore such data spaces.
\end{RQ}

\section{Threats to Validity} \label{threats}
    \subsection{Sampling Bias}
    This paper shares the same sampling bias problem as every other data mining paper. Sampling bias threatens any classification experiment; what matters in one case may or may not hold in another case.  For example, even though we use 10 open-source datasets in this study which come from several sources, they were all supplied by individuals. 
    
    As researchers, we can adopt two tactics to reduce the sampling bias problem. First   we can     document our tools and methods, then post an executable  reproduction package for all the experiments (that package for this paper is available at  \url{url_blind_for_review}). 
    
    Secondly, when new data becomes available, we can test our methods on the new data. For example, \tabl{issue_lifetime} shows
    results were FFTs and four different state-of-the-art learners, i.e. Decision Tree, Random Forest, Logistic Regression, K-Nearest Neighbors, were applied
    to the task of predicting issue close time (the other four learners were 
    used since that was the technology recommended in a recent
    study in that domain~\cite{rahul2018bellwhether,rees2017better}). Unlike the
    defect prediction data, we did not have multiple versions of the code so, for this domain, we used a 5*10-way cross-validation
    analysis. White cells show where the FFT results were statistically different and better than all of the state-of-the-art learners' results.
    Note that, in  most cases  ($43/56=77\%$),  
    FFTs performed better. 
    
    While this result does not prove that FFTs works well
    in all domains, it does show that there exists more than one
    domain where this is a useful approach.

\begin{table}
\centering
\caption{Which learners performed better (in terms of   median {\em Dis2heaven}) in  5*10 cross-value experiments predicting
for different classes of ``how long to close an Github issue''.
Gray areas denote experiments where FFTs were out-performed by other learners.  Note that, in (43/56=77\%) experiments, FFT performed better
than the prior state-of-the-art in this area~\cite{rahul2018bellwhether}.}
\label{tab:issue_lifetime}
\resizebox{0.45\textwidth}{!}{%
\begin{threeparttable}
\begin{tabular}{|l|l|l|l|l|l|l|l|}
\hline
                                & \multicolumn{7}{c|}{\textbf{Days till closed}}                                                                                                                                                                 \\ \cline{2-8} 
\textbf{Data(\# of instances)} & \textbf{$>365$ }            & \textbf{$<180$  }            & \textbf{$<90$  }                      & \textbf{$<30$  } & \textbf{$<14$  } & \textbf{$<7$  }              & \textbf{$<1$  }               \\ \hline
cloudstack (1551)               & FFT                            & FFT                           & FFT                                 & FFT                   & FFT                   & \cellcolor[HTML]{C0C0C0}DT & \cellcolor[HTML]{C0C0C0}LR \\ \hline
node (6207)                     & FFT                            & FFT                           & FFT                                 & FFT                   & FFT                   & \cellcolor[HTML]{C0C0C0}DT & \cellcolor[HTML]{C0C0C0}LR \\ \hline
deeplearning (1434)             & FFT                            & FFT                           & FFT                                 & FFT                   & FFT                   & FFT                        & \cellcolor[HTML]{C0C0C0}RF \\ \hline
cocoon (2045)                   & FFT                            & FFT                           & FFT                                 & FFT                   & FFT                   & FFT                        & FFT                        \\ \hline
ofbiz (6177)                    & FFT                            & FFT                           & FFT                                 & FFT                   & FFT                   & FFT                        & FFT                        \\ \hline
camel (5056)                    & \cellcolor[HTML]{C0C0C0}RF/KNN & \cellcolor[HTML]{C0C0C0}KNN   & \cellcolor[HTML]{FFFFFF}FFT/KNN/DT  & FFT                   & FFT                   & FFT                        & FFT                        \\ \hline
hadoop (12191)                  & \cellcolor[HTML]{C0C0C0}KNN    & \cellcolor[HTML]{C0C0C0}DT    & \cellcolor[HTML]{C0C0C0}\textit{DT} & FFT                   & FFT                   & FFT                        & FFT                        \\ \hline
qpid (5475)                     & \cellcolor[HTML]{C0C0C0}DT     & \cellcolor[HTML]{C0C0C0}DT/RF & \cellcolor[HTML]{C0C0C0}DT          & FFT                   & FFT                   & FFT                        & FFT                        \\ \hline
\end{tabular}
\begin{tablenotes}
    \item The goal here is to classify an issue according to how long it will
take to close; i.e. less than  1 day, less than 7 days, and so on. 
Values collected via a 5x10 cross-validation procedure.   
   \item Cells with a (white, gray) background means FFTs are statistically (better, worse) than (all, any) of the state-of-the-art learners (as determined  by a Mann-Whitney test, 95\% confidence), respectively. KNN, DT, RF and LR represents K-Nearest Neighbors, Decision Tree, Random Forest and Logistic Regression respectively.
\end{tablenotes}
\end{threeparttable}
}
\end{table}
    \subsection{Learner Bias}
        For building the defect predictors in this study, we elected to use Simple Logistic, Naive Bayes, Expectation Maximization, Support Vector Machine. We chose these learners because past studies shows that, for defect prediction tasks, these four learners represents four different levels of performance among a bunch of different learners~\cite{ghotra2015revisiting,agrawal2018better}. Thus they are selected as the state-of-the-art learns to be compared with FFTs on the defect prediction data. While for \tabl{issue_lifetime}), K-Nearest Neighbors, Decision Tree, Random Forest and Logistic Regression are used to compare against FFTs, because a recent work has summarized all the best learners that were applied on the issue lifetime data.

    \subsection{Evaluation Bias}
        This paper uses two performance measures, i.e., $P_{opt}$ and $dist2heaven$ as defined in Equation \ref{eq:popt} and \ref{eq:d2h}. Other quality measures often
        used in software engineering to quantify the effectiveness of prediction ~\cite{menzies2007problems, menzies2005simple, jorgensen2004realism}. A comprehensive analysis using these measures may be performed with our replication package. Additionally, other measures can easily be added to extend this replication package.
    
    \subsection{Order Bias}
    For the performance evaluation part, the order that
    the data trained and predicted affects the results.
    
    For the defect prediction datasets,
    we deliberately choose an ordering that mimics
    how our software projects releases versions so, for those experiments, we would say that bias was a required and needed.
    
    For the issue close time results of \tabl{issue_lifetime}, to 
      mitigate this order bias, we ran our rig in a  the 5-bin cross validation 10 times, randomly changing the order of the data each time.

\section{Conclusions} \label{conclusions}

This paper has shown that a data mining algorithm 
call Fast-and-Frugal trees (FFTs) developed
by psychological scientist is remarkably effective for creating actionable software
analytics. Here ``actionable'' was defined as a
combination of comprehensible and operational.

Measured in terms of comprehensibility, the
FFT examples of Table~\ref{tab:three} show that
FFTs satisfy requirements raised by psychological scientists for ``easily understandable at an expert level''; i.e., they comprise several short rules and those rules can be quickly applied (recall that each level of an FFT has an exit point which, if used, means humans can ignore the rest of the tree).

Despite their brevity, FFTs are remarkably effective:
\bi
\item
Measured in terms of $P_{\mathit{opt}}$, FFTs are much better than
other standard algorithms (see \fig{cmp-dt2}).
\item
Measured in terms of distance to the ``heaven'' point of 
100\%  recall and no false alarms, FFTs are either usually better  than other standard algorithms used in software analytics (Random Forests, Naive Bayes, EM, Logistic Regression, and SVM). This result holds for
at least two SE domains: defect prediction (see Figure~\ref{fig:cmp-dt2})
issue close time prediction (see \tabl{issue_lifetime}). 

\ei
As to being operational,  we found that if learning is restricted to just the attributes changed most often, then the behavior of other learning algorithms can vary, wildly (see \fig{diffs}). The behaviour of FFTs, on the other hand, remain remarkable stable across that treatment.

From the above, our conclusions is two-fold:
\begin{enumerate}
\item
There is much the software analytics community could learn from psychological science. FFTs, based on psychological science principles, 
out-perform a wide range of  learners in widespread use.
\item
Proponents of complex methods should always baseline those methods against simpler alternatives. For example, FFTs could be used as a standard baseline learner against which other software  analytics tools are compared.
\end{enumerate}

\section{Future Work}

Numerous aspects of the above motivate deserve more
attention.

\subsection{ More Data}
This experiment with issue close time shows that FFTs are useful for more just defect prediction data. That said, for future work, it is important to test many other SE domains to learn when FFTs are useful. 
For example, at this time we are exploring text
mining of StackOverflow data.

\subsection{  More Learners} The above experiments
should be repeated, comparing FFTs against more learners.
For example, at this time, we are comparing FFTs
against deep learning for SE  datasets. At this time,  there is nothing
as yet definitive to report about those results.

\subsection{ More Algorithm Design}
These results may have implications beyond SE.
Indeed, it might be insightful to another field-- machine learning.
For the reader familiar with machine learning literature,
we note that FFTs are a decision-list rule-covering model.  FFTs restrict the (a)~number of conditions per rule to only one comparison and
(b)~the total number of rules is set to a small number (often often just $d\in \{3,4,5\}$).
Other decision list approaches such as
PRISM~\cite{cendrowska1987prism}, INDUCT~\cite{Witten:2002},RIPPER~\cite{Cohen95fasteffective} and
RIPPLE-DOWN-RULES~\cite{Gaines:95}  produce far more complex models since they impose no such restriction.
Perhaps the lesson of FFT is that PRISM,INDUCT,RIPPER, etc could be simplified with a few simple restrictions on the models they learn.

Also  the success of FFT might be credited to
its use on ensemble methods; i.e. train multiple times,
then select the best. 
The comparison between FFTs and other ensemble methods 
like bagging and boosting~\cite{quinlan1996bagging} could be useful in future work.

\subsection{ Applications to Delta Debugging}
There is a potential connection between
the \fig{diffs}  results and the delta debugging results of Zeller~\cite{Zeller:2002}.
 As shown above, we  found that, sometimes
 focusing on the values that change most can sometimes, lead to better defect predictors (though, caveat empty or, sometimes it can actually make matters worse-- see the large negative results in \fig{diffs}). Note that this parallels Zeller's approach which
 he summarizes as  
``Initially, variable v1 was x1, thus variable v2 became x2, thus variable v3 became x3 ... and thus the program failed''.   In future work, we
will explore further applications of FFTs to delta debugging.

\balance
\bibliographystyle{ACM-Reference-Format}

\end{document}